# Effect of doping on the linear temperature dependence of the magnetic penetration depth in cuprate supraconductors

J. Le Cochec[a], G. Lamura[a], A. Gauzzi[b*], F. Licci[b], A. Revcolevschi[c], A. Erb[d], G. Deutscher[e] and J. Bok[a].

[a]Laboratoire de Physique du Solide, ESPCI, 10, rue Vauquelin, F-75231 Paris, France.
[b]MASPEC-CNR Institute, Area delle Scienze, 43010 Parma-Fontanini, Italy.
[c]Laboratoire de Physico-chimie des Solides, UMR 8648, Univ. Paris Sud, 91405 Orsay, France.
[d]DPMC, Université de Genève, 1200 Genève, Switzerland.
[e]School of Physics and Astronomy, Tel Aviv University, Ramat Aviv, Tel Aviv 69978, Israel.

We report on measurements of the low temperature dependence of the magnetic penetration depth $\lambda_{ab}$ in several single crystals of $La_{2-x}Sr_xCuO_4$, $Bi_2Sr_2CaCu_2O_8$ and $YBa_2Cu_3O_{6+x}$ at various doping levels ranging from the under- to the overdoped regimes by using a novel single-coil technique achieving 10 pm resolution. We have found a linear dependence of $\lambda_{ab}$ in all samples with a rapidly increasing slope $d\lambda_{ab}/dT$ as doping decreases. Our analysis of the data indicates that a superconducting gap with *d*-wave symmetry is sufficient to quantitatively account for the above slope values in the optimally or over- doped samples. In the underdoped samples, the *d*-wave model predicts much smaller values than those measured by assuming realistic values for the zero-temperature $\lambda_{ab}$ and gap $\Delta$. The experimental data are compatible with a model of thermodynamic phase fluctuations of the order parameter. Therefore, we put forward the hypothesis that the gapless properties observed in cuprates may have qualitatively different physical origins depending on the doping level.

## 1. INTRODUCTION

It has been established that most cuprate superconductors display gapless properties at low temperature. To account for these properties, the existence of lines of nodes in the gap function, as in the case of a *d*-wave gap, is currently invoked [1,2]. Alternative explanations exist, such as those based on thermally activated phase fluctuations of the gap [3]. This picture was already proposed back in the 70's for granular superconductors [4] and is suited for metals with low density of carriers and reduced dimensionality, which is indeed the case of cuprates. In order to establish whether such fluctuations are relevant or not in cuprates, we report here on measurements of the variations of the magnetic penetration depth $\lambda_{ab}$ at different doping levels, since these variations are directly related to the low energy excitation spectrum.

___

* To whom correspondence should be addressed:
E-mail : gauzzi@maspec.bo.cnr.it

## 2. EXPERIMENTAL

We have performed our measurements on four single crystals: one underdoped $La_{1.86}Sr_{0.14}CuO_4$ (LSCO) [5] with $T_c$=24 K, one optimally doped $YBa_2Cu_3O_{6.9}$ (YBCO) [6] ($T_c$ = 91 K) and one underdoped $Bi_2Sr_2CaCu_2O_8$ (BSCCO) [7] ($T_c$=83 K). The BSCCO sample was subsequently overdoped by annealing it at 750 °C in flowing oxygen for 84 hours. The resulting $T_c$ after this treatment was 61 K. The $\lambda_{ab}$ measurements at low temperature have been performed with a novel single coil mutual inductance technique with very high resolution in the 10 pm range, as described elsewhere [8].

## 3. RESULTS AND DISCUSSION

In figs. 1 and 2, we report the low temperature dependence of $\lambda$ measured in the above samples. In Tab. 1 we report the experimental values of

$d\lambda_{ab}/dT$ and those predicted by the above two models: 1) simple *d*-wave; 2) phase fluctuations. We recall that the prediction of the first model requires the knowledge of $\lambda_{ab}(0)$ and of the zero-temperature superconducting gap $\Delta_0$ (not the pseudogap). As to the prediction of 2), we have extended the validity of the formula proposed by Roddick and Stroud [3] to the more general case of anisotropic continuous medium, which is suited to describe cuprates. Within the Gaussian approximation, the result is the following:

$$\frac{d\lambda_{ab}(T)}{dT} \approx \frac{\mu_0 k_B}{\Phi_0^2} \lambda_{ab}^3(0) \left(\frac{m_c^*}{m_{ab}^* \xi_{ab}^2 \xi_c}\right)^{\frac{1}{3}} \quad (1)$$

where $\Phi_0$ is the flux quantum, the indices *ab* and *c* refer to the *ab*-plane and *c*-axis respectively, $m^*$ is the effective mass and $\xi$ is the BCS coherence length. Eq. (1) requires the knowledge of a second additional parameter, which is the quantity in parentheses. Neither $\lambda_{ab}(0)$ nor such second quantity are directly measured in our case, so we have borrowed realistic parameters from the literature and obtained the values reported in Tab. 1.

We estimate the uncertainty on both parameters to be less than 100%, then the large differences found in Tab. 1 between experimental and predicted values allow us to conclude the following. 1) the *d*-wave symmetry quantitatively accounts for the experimental values observed in the optimally doped YBCO and overdoped BSCCO; 2) both underdoped samples exhibit much larger values than those predicted by the *d*-wave model for any realistic values of $\lambda_{ab}(0)$ and $\Delta_0$. Such large slopes are compatible with the phase fluctuation model leading to eq. (1).

This result suggests that the gapless properties observed in cuprates may have different origins depending on the doping level. A more extensive study on more samples in both under- and overdoped regimes is required to confirm this conclusion. Nevertheless, the high uniformity of the BSCCO samples rules out the possibility that extrinsic factors may be responsible for the large slope value observed in the underdoped regime.

We acknowledge T. Besagni and P. Ferro for their help in sample preparation.

Table 1. Experimental and theoretically expected values for $d\lambda_{ab}/dT$. Values of $\Delta_0$ calculated from $2\Delta_0 = 5 k_B T_c$. Units are Å and K. $\lambda_{ab}$ values used for the d-wave prediction are also shown.

| Samples | Exp. | $\lambda_{ab}(0)$ | d-wave | Phase fluct. |
|---|---|---|---|---|
| YBCO | 4.5±0.1 | 1500 | 4.5 | 0.5 |
| LSCO | 167.5±0.9 | 6500 | 98 | 160 |
| BSCCO-under | 42.4±0.2 | 2500 | 10 | 40 |
| BSCCO-over. | 10.3±0.1 | 1800 | 10 | 4 |

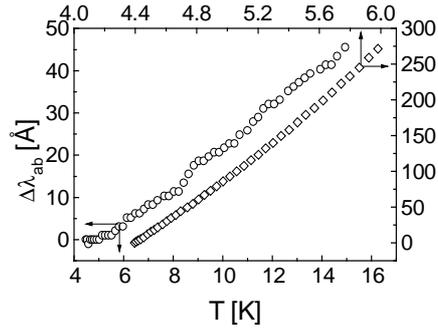

Fig. 1. Low temperature penetration depth for the underdoped LSCO (diamonds) and optimally doped YBCO (circles). Note the two different scales

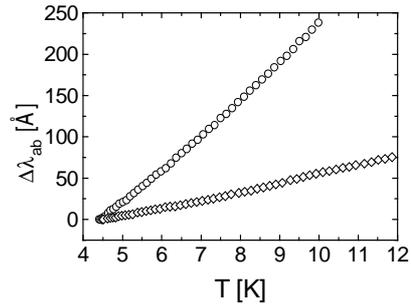

Fig. 2. The same as in fig. 1 for the underdoped (circles) and overdoped (diamonds) BSCCO samples.

**REFERENCES**

1. W. Hardy *et al.*, Phys. Rev. Lett. **70**, 3999 (1993).
2. S. Lee *et al.*, Phys. Rev. Lett. **77**, 735 (1996).
3. E. Roddick and D. Stroud, Phys. Rev. Lett. **74**, 1430, (1995).
4. G. Deutscher *et al.*, Phys. Rev. B **10**, 4598 (1974).
5. A. Revcolevschi and J. Jegoudez in *Coherence Effects in High Temperature Superconductors* (World Scientific, Singapore, 1996), p. 19.
6. A. Erb *et al.*, Physica C **245**, 245 (1995).
7. F. Licci, unpublished.
8. A. Gauzzi *et al.*, preprint.